\documentclass{article}
\usepackage{epsfig}

\date{\today}

\usepackage{amsmath}
\usepackage{amssymb}
\usepackage[hang,nooneline,scriptsize]{subfigure}
\begin{document}

\title{Gravitating Q-tubes and cylindrical spacetime}

\author{{\large Yves Brihaye \footnote{email:
yves.brihaye@umons.ac.be}}$^{\dagger}$ and
  {\large T\'erence Delsate \footnote{email:
terence.delsate@umons.ac.be}}$^{\dagger}$
\\ \\
$^{\dagger}${\small Physique-Math\'ematique, Universit\'e de
Mons, 7000 Mons, Belgium}\\ 
}

\date{\today}
\setlength{\footnotesep}{0.5\footnotesep}
\newcommand{\dd}{\mbox{d}}
\newcommand{\tr}{\mbox{tr}}
\newcommand{\la}{\lambda}
\newcommand{\ka}{\kappa}
\newcommand{\f}{\phi}
\newcommand{\vf}{\varphi}
\newcommand{\F}{\Phi}
\newcommand{\al}{\alpha}
\newcommand{\ga}{\gamma}
\newcommand{\de}{\delta}
\newcommand{\si}{\sigma}
\newcommand{\bomega}{\mbox{\boldmath $\omega$}}
\newcommand{\bsi}{\mbox{\boldmath $\sigma$}}
\newcommand{\bchi}{\mbox{\boldmath $\chi$}}
\newcommand{\bal}{\mbox{\boldmath $\alpha$}}
\newcommand{\bpsi}{\mbox{\boldmath $\psi$}}
\newcommand{\brho}{\mbox{\boldmath $\varrho$}}
\newcommand{\beps}{\mbox{\boldmath $\varepsilon$}}
\newcommand{\bxi}{\mbox{\boldmath $\xi$}}
\newcommand{\bbeta}{\mbox{\boldmath $\beta$}}
\newcommand{\ee}{\end{equation}}
\newcommand{\eea}{\end{eqnarray}}
\newcommand{\be}{\begin{equation}}
\newcommand{\bea}{\begin{eqnarray}}

\newcommand{\ii}{\mbox{i}}
\newcommand{\e}{\mbox{e}}
\newcommand{\pa}{\partial}
\newcommand{\Om}{\Omega}
\newcommand{\vep}{\varepsilon}
\newcommand{\bfph}{{\bf \phi}}
\newcommand{\lm}{\lambda}
\def\theequation{\arabic{equation}}
\renewcommand{\thefootnote}{\fnsymbol{footnote}}
\newcommand{\re}[1]{(\ref{#1})}
\newcommand{\R}{{\rm I \hspace{-0.52ex} R}}
\newcommand{\N}{{\sf N\hspace*{-1.0ex}\rule{0.15ex}%
{1.3ex}\hspace*{1.0ex}}}
\newcommand{\Q}{{\sf Q\hspace*{-1.1ex}\rule{0.15ex}%
{1.5ex}\hspace*{1.1ex}}}
\newcommand{\C}{{\sf C\hspace*{-0.9ex}\rule{0.15ex}%
{1.3ex}\hspace*{0.9ex}}}
\newcommand{\eins}{1\hspace{-0.56ex}{\rm I}}
\renewcommand{\thefootnote}{\arabic{footnote}}
 \maketitle
\begin{abstract} 
We consider a model involving a self-interacting complex scalar field minimally 
coupled to gravity and emphasize the cylindrically symmetric classical solutions. 
A general ansatz is performed which transforms the field equations
into a system of differential equations.  In the generic case, the scalar field
depends on the four space-time coordinates.
The underlying Einstein vacuum equations lead to a family of explicit solutions  
 extending the Kasner space-time. 
The solutions of the coupled system are -static as well as stationnary- gravitating Q-tubes
of scalar matter which deform space-time.
\end{abstract}
\medskip 
\medskip
 \ \ \ PACS Numbers: 04.70.-s,  04.50.Gh, 11.25.Tq

\section{Introduction}
Nontopological solitons are known for a long time since the pioneering work of
Friedberg, Lee and Sirlin \cite{friedberg}, see also  \cite{lee_pang} for a review.
Among the most studied of this type of classical solutions are  Q-balls \cite{coleman}.
Such solitons are classical objects build out of scalar field, they are localized in space and stabilized due to a conserved charge
associated to an U(1) global symmetry of the underlying field theory. They  occur
in several contexts of particle theory; for example in the supersymmetric extensions of the
Standard Model \cite{kusenko}. Possibilities that Q-balls could play a role in the baryon asymmetry
of the Universe and dark matter were emphasized, for example, in \cite{bau} and \cite{dark} respectively. 

While the Q-balls are generically seen as spherically symmetric objects, the classical equations can
admit solutions with different types of symmetries and topology, for example solutions with a cylindrical shape
(or string shape) can bee looked for \cite{Sakai:2010ny}~: the so called Q-tubes. These solutions
can be seen as two-dimensionnal Q-balls embedded in three-dimensional space time. An unified picture
of Q-balls and Q-tube  is emphasized in \cite{Tamaki:2012yk}. 
The Q-tube solutions can be made spinning and were first constructed in \cite{Volkov:2002aj} where the
question of spinning Q-balls was adressed in two and three dimensional space-time.

When set in an appropriate context, the Q-balls and Q-tubes can be very large
and massive.  Accordingly they can be of astronomical  size and it  becomes
necessary to take the gravitational effect into account. The gravitating Q-balls
are called boson stars, the patterns of solutions then become very rich and 
several features of boson stars were reported e.g. in \cite{kkl}.  Further properties
of gravitating Q-balls were discussed in \cite{Tamaki:2011bx}. 

In this paper we address the construction of gravitating Q-tubes including a
rotation in the plane perpendicular to the tube and a momentum (or boost) along the axis.
We first discuss the metric that incorporates these two effects. The spatial
part of the metric describes a rotating-boosted cylinder geometry. It turns out that
 analytic results can be obtained for the vacuum Einstein equations \cite{clement}. 

For the matter fields, we consider  a complex scalar field interacting through 
an appropriate U(1)-invariant potential.
The corresponding  globally symmetric lagrangian  is  coupled minimally to gravity. 
To implement the cylindrical symmetry,  we use for the fields an extension of the ansatz 
used in \cite{Verbin:1998tc,Christensen:1999wb}.
Non-linear sigma models analogous to the model under investigation  were studied namely in \cite{Verbin:2004er}. 

Solving the full equations bu means of a numerical technique,
we manage to construct several families of Q-tubes and we argue that they
exist in some range of the parameters encoding their spin or boost. However
 it turns out rather difficult to construct Q-tubes in the generic case,
 i.e. with both rotation and boost. 

Let us finally point out that the model  under consideration in this paper is a particular case of the more involved
superconducting string model \cite{p_peter} which also contains U(1) a gauge symmetry and a second complex scalar field. 
The equations studied in the present paper correspond 
to a particular limit of  the ones solved in \cite{Hartmann:2012mh}.
The particular case corresponding to gravitating Q-tubes and the
incorporation of the rotations was, however, not emphasized by these authors.

\section{Model and ansatz}
We consider a complex  scalar field $\phi(x)$ self-interacting through a 
specific U(1)-invariant potential and minimally coupled to gravity.
The corresponding action reads~:
\begin{equation}
S= \frac{1}{16\pi G} \int d^4 x \sqrt{-g}  
\left( R + 16\pi G {\cal L}_{\rm matter} \right) \ , 
\end{equation}
where ${\cal L}_{\rm matter}$ denotes the matter Lagrangian~:
\begin{equation}
\label{lmatter}
{\cal L}_{\rm matter}= \frac{1}{2}
\left(\partial_{\mu} \phi \right)^* \partial^{\mu} \phi - V(|\phi|)  \ \ , V(z) = m^2 (\frac{1}{2} z^2 - \gamma z^4 + \kappa z^6) .
\end{equation}
Here we employ for the  potential the form used in \cite{Volkov:2002aj} (other forms are also used
currently, see e.g. \cite{Tamaki:2012yk}).  
It is well known that the above action possesses a Noether current 
$j_{\mu} = -i(\phi^* \partial_{\mu} \phi - (\partial_{\mu} \phi^*) \phi)$.  
The corresponding  conserved charge is noted $Q$.

The coupled matter and gravity  field equations are obtained from the variation of the
action with respect to the scalar and metric fields, respectively~:
\begin{equation}
 \nabla^{\mu} \nabla_{\mu} \phi = - m^2 \phi (1 - 4\gamma |\phi|^2 + 6 \kappa |\phi|^4) \ \ ; 
 \ \  G_{\mu \nu} =8\pi G T_{\mu \nu } \ ,  \ \mu, \nu=0,1,2,3, \ ,
\end{equation}
and where  $T_{\mu \nu}$ is the energy-momentum tensor
\begin{equation}
 T_{\mu \nu}=-(g_{\mu \nu } {\cal L}_{\rm matter} - 2\frac{\partial {\cal L}_{\rm matter}}{\partial g^{\mu \nu}})
 = \frac{1}{2} \left(\partial_{\mu} \phi ^* \partial_{\nu} \phi + \partial_{\nu} \phi^*  \partial_{\mu} \phi \right)
 - g_{\mu \nu} {\cal L}_{\rm matter} \ .
\end{equation}
We want to study regular solitons of the above equations presenting a cylindrical symmetry about the axis $z$. 
In general, such  solutions a characterized by one angular momentum and
eventually a linear momentum about the $z$ axis, a boost. The most general metric compatible
with the above requirements and the cylindrical symmetry can be parametrized according to  
\begin{eqnarray}
ds^2 &=&  g_{ab}(R) dx^a dx^b - dR^2 \\
     &=&  N^2 dt^2 - dR^2 - L^2 (d \varphi - W dt) ^2 - K^2 (dz - M d\varphi - v dt) ^2 ,
\label{metric}
\end{eqnarray}
where $R$ denotes the radial variable in the $x,y$-plane,
$\varphi$ the corresponding angle (varying from $0$ to $2 \pi$) and the
functions $N,L,K$ and $W,M,v$ depend on $R$ only.
This metric is such that $\sqrt{-g} = KLN$; it reduces to the
ansatz used in \cite{Verbin:1998tc} in the limit $M=v=0$.

We complete the ansatz by choosing an harmonic dependance on $\varphi$ and $z$
for the complex scalar field:
\begin{equation}
  \phi(x) = f(R) \exp(i(\omega t + n \varphi + \lambda z)),
\end{equation}
 where $\omega,\lambda\in\mathbb R,\ n\in \mathbb N$.
Note that the scalar field depends on \emph{all} the spacetime variables. The
form of the scalar field is such that the explicit dependance on the
$z,t,\varphi$ variables disappears from the equations.
We will refer to the case where $\lambda \neq 0$ as boosted and the case where
$n \neq 0$ as rotating.
 
With the ansatz above, the scalar field equation  reduces to the differential equation
\be
 f'' + (\frac{N'}{N}+\frac{L'}{L}+\frac{K'}{K} ) f' +\frac{1}{N^2}
(W(\lambda M+n)+\lambda  v+\omega )^2f
-\frac{1}{L^2}(\lambda M+n)^2f - \frac{\lambda^2}{K^2}f= \frac{d}{df}V(f),
\ee
where the prime here denotes the derivative with respect to $R$.
 The resulting
Einstein equations (which we do not write  explicitely for shortness)
can easily be reconstructed from the identities presented in the next section
 and in Appendix A1. 

In this paper, we mainly focus the discussion to the spinning case; 
that the boosted case is qualitatively similar.
In thes two particular cases, the function $M$ vanishes
identically.  This function is non-trivial only in the generic case where both rotation and
boost are present. The equations for the general system are given in Appendix
\ref{app:eqs}. 
It turns out that switching on both the boost and the rotation 
leads
to a delicate numerical problem. We were  able to integrate the full system in
this case, but the interpretation of the resulting solutions is still unclear.
However, as we will further argue, such solutions must exist and consistently
solve the Einstein equations and the matter field equation.

\section{Vacuum solutions}
We are interested in solutions where the matter field is localized around the
symmetry axis of the tube. In this respect, the knowledge of the vacuum
solutions  is usefull.

In the non-rotating and non-boosted case $W=0, v=0$, it is well known that the
solutions are given by the family of  'Kasner-space-time' 
labelled by the Kasner parameters $a,b,c$~:
\begin{equation}
ds^2 =  \gamma_t R^a dt^2 - dR^2 - (1+ \Delta) R^b d \varphi^2 - \gamma_z R^c
dz^2
\end{equation}
with the conditions  $a+b+c = 1$, $a^2+b^2+c^2 = 1$. 
The parameter $\Delta$ characterizes the angular deficit of the solution; 
$\gamma_t$ and $\gamma_z$ are not intrisic~: they 
can be rescaled in the variables $t$ and $z$ respectively.

Solving the coupled equations (i.e. with matter) leads to specific
values of
the parameters $a,b,c$ and $\Delta$,  depending of the coupling constants of
the
potential and of $\omega$.

\subsection{Integrals of motion}
In order to discuss the rotating or boosted case, it is useful to note that six
out of the seven independant Einstein vacuum equations  lead to first integrals.
Indeed, the relevant components of the Ricci tensor can be set in the form 
\bea
&&R_{t}{}^{t} =-\frac{K_t'}{2\sqrt{-g}},\ R_{t}{}^{\varphi}
=-\frac{K_\varphi'}{2\sqrt{-g}},\ R_{t}{}^{z}
=-\frac{K_z'}{2\sqrt{-g}},\nonumber\\
&&R_\varphi{}^\varphi = -\frac{K_{\varphi\varphi}'}{2\sqrt{-g}},\ R_\varphi{}^z
=-\frac{K_{\varphi z}}{2\sqrt{-g}},\ R_z{}^z = \frac{K_{zz}}{2\sqrt{-g}},
\eea
where
\bea
K_t &=& \frac{K^3 L v W M'}{N}-2K L N'+\frac{K^3 L vv'}{N}+\frac{KL^3 W
W'}{N},\nonumber\\
K_\varphi &=& \frac{K^3 LM WM'}{N}+\frac{K^3 LMv'}{N}-\frac{K L^3
W'}{N},\nonumber\\
K_z &=& \frac{K^3 L \left(W M'+v'\right)}{N},\nonumber\\
K_{\varphi\varphi}&=& - K N L'+\frac{K^3 LM W^2 M'}{N}-\frac{K^3 MN
   M'}{L}+\frac{K^3 L M W v'}{N}-\frac{K L^3 W W'}{N},\nonumber\\
K_{\varphi z} &=&\frac{K^3 L W^2M'}{N}-\frac{K^3 N M'}{L}+\frac{K^3 L W
v'}{N},\nonumber\\
K_{zz}&=&-2 LNK'-v \left(\frac{K^3 LWM'}{N}+\frac{K^3Lv'}{N}\right)
-M\left(\frac{K^3 M' \left(L^2W^2-N^2\right)}{LN}+\frac{K^3 L W
v'}{N}\right).
\label{firstint}
\eea

We further note that the remaining equation $R_r{}^r = 0$ reduces to
\be
R_r{}^r = \frac{K''}{K}+\frac{L''}{L}+\frac{N''}{N} + \frac{K_1 + 2 L_k (K
N)'}{2 K^4 M_k^2 N} -\frac{  (L_k^2 + K^2M_k^2)}{K^4 L^4},
\ee
where we defined 
\be
K_1 = K_t K_z + K_z K_{zz} + K_\varphi K_{\varphi z},M_k= K_\varphi- M K_z,\
L_k= K_z L.
\ee

A suitable combination of (\ref{firstint}) leads to
\be
(NLK)' = -\frac{1}{2}(K_{zz}+K_{\varphi \varphi} + K_t),
\ee
which imposes that in the general case, at least half of the Kasner relations
hold ( $a+b+c=1$, assuming $N,(K,L)\propto r^a,(r^b,r^c)$ resp.)

\subsection{Solutions in matrix form}
 Using the formalism developped in \cite{clement}, it is possible to
find the general solution of the system $R_{\mu}^{\nu}=0$ in a closed form.
In fact the solutions of vacuum-Einstein equations corresponding to the metric 
(\ref{metric}) can be written  in the following form involving two  3$\times$3 matrices $A$ and $C$~:
\be
     g(R) = C \exp [2 A \ln (R-R_0)] \ \ , \ \ {\rm Tr} A = {\rm Tr} A^2 = 1 \ \ , \ \ C^T = C \ \ , \ \  (C A)^T = CA.
\ee
The three eigenvalues of $A$, say $a,b,c$, are the Kasner powers mentionned above.
Assuming these eigenvalues  to be non-degenerate, we can further write
\be
     g(R) = C {\cal A} \ {\rm diag}( (R-R_0)^{2a} , (R-R_0)^{2b}, (R-R_0)^{2c} ) \ {\cal A}^{-1} \ \ , \ \ 
     A =  {\cal A} \ {\rm diag} (a,b,c) \ {\cal A}^{-1}
\ee 
The extraction of the  functions parametrizing the metric (\ref{metric}) leads in general to cumbesome
rationnal functions of $R$. 

\subsection{Rotating case}
In the case $W \neq 0,\ v=M=0$, the expression for $g$ simplifies considerably.
If we further require the metric to be a deformation of the Minkowski space-time,
i.e. with $ |b-1| \sim 0$, $a \sim 0$, $c \sim 0$ the matrices ${\cal A},C$ need to be of special form. We find
\be
\label{aparticulier}
       {\cal A} = \left( \begin{array}{ccc}
                                  1 &-\frac{c_{12}}{c{11}}&0  \\
                                  0 &1            &0  \\
                                  0 &0            &1  \\
                                            \end{array}\right) 
\ee
leading to
\be
\label{gparticulier}
 g = \left( \begin{array}{ccc}
               c_{11} R^{2a} &c_{12}R^{2a}&0 \\
               c_{12} R^{2a} &\frac{(c_{11}c_{22} - c_{12}^2)}{c_{11}}R^{2b} + \frac{c_{12}}{c_{11}}^2R^{2a}   &0 \\
                0 &0& c_{33} R^{2c} \\
                                \end{array}   \right) 
\ee
with obvious definitions for $c_{ij}$ and  $c_{13}=c_{23}=0$.

The  diagonal 'main' functions $N,L,K$ obey the Kasner form only asymptotically and 
the following exact relations hold
\be
\label{asymptotic_w}
      W' = -K_\varphi \frac{N}{K L^3} \ \ \ , \ \ \  K = K_0 (N L)^{c/(a+b)} \ \ ,
\ee
with  the ensuing asymptotic form for  the function $W$ and  the non-diagonal element of the
metric~: 
\be
         \ \ W(R) \sim -\frac{K_\varphi}{R^{2(b-a)}} \ , g_{0 2}\ = L^2 W
\sim W_0 R^{2a}
\ee 
As long as the Kasner powers $b,c$ are such that $2b+c > 1$, the 'W-terms' in
the equations are subdominant with respect to terms involving  the diagonal function.  
The solutions that we will report in the next section are mainly of this type.

\subsection{Boosted case}
In the case where $v\neq0,\ W=M=0$, 
the result of the previous subsection can be adapted by exchanging the role of 
the coordinates $\varphi$ and of $z$. The relevant integral of motion leads to 
\be
v' = K_z \frac{N}{K^3L} \ \ , \ \  v(r) \sim \frac{K_z}{R^{2(c-a)}},
\ee

The generalisation of the Eq. (\ref{aparticulier}) to the case with boost and rotation
is  obtained by choosing for $C$ an arbitrary symmetric matrix anf for $\cal A$
 an upper triangular matrix of the form
\be     
       {\cal A} = \left( \begin{array}{ccc}
                                  1 &A_{12}&A_{13}  \\
                                  0 &1            &A_{23}  \\
                                  0 &0            &1  \\
                                            \end{array}\right) 
\ee
where
$$
A_{12} =  -\frac{c_{12}}{c_{11}} \ , \ 
A_{13} = \frac{c_{12}c_{23} -  c_{12}c_{22}}{c_{11}c_{22}- c_{12}^2}             \ , \ 
A_{23} = \frac{c_{12}c_{13} -  c_{11}c_{23}}{c_{11}c_{22}- c_{12}^2} 
$$
As we will point out in the next section, in the case of spinning, boosted
solutions with matter fields the asymptotic metric is not compatible  
with the form for generic values of the parameters $\omega, \lambda$. 
This suggests that only a very small range
of parameters lead to boosted and rotating configurations where both Kasner conditions
are fulfilled. This deserves further investigations.

From now on, we focus on the rotating case only except if stated explicitly.

\section{Equations  and physical quantities}
\subsection{Energy momentum tensor}
From now on, we concentrate on the case with no boost (i.e. $M=v=0$).
In order to establish the Einstein equations, we need the energy momentum tensor associated with the scalar field.
The non vanishing  components of this tensor  read~:
\begin{eqnarray}
T_t^t &=& V(f) + \omega^2 \frac{f^2}{2 N^2} + \frac{(f')^2}{2} + f^2 n^2 \frac{N^2-L^2 W^2}{2 N^2 L^2} \ \ , \ \ \nonumber \\
T_r^r &=& V(f) - \omega^2 \frac{f^2}{2 N^2}  - \frac{\omega n f^2}{N^2} - \frac{(f')^2}{2} + f^2 n^2 \frac{N^2-L^2 W^2}{2 N^2 L^2}
   \ \ , \ \ \nonumber \\
T_{\varphi}^{\varphi} &=& V(f) - \omega^2 \frac{f^2}{2 N^2} + \frac{(f')^2}{2}- f^2 n^2 \frac{N^2-L^2 W^2}{2 N^2 L^2}
 , \ \ \nonumber \\
T_z^z &=& V(f) - \omega^2 \frac{f^2}{2 N^2}  - \frac{\omega n f^2}{N^2} + \frac{(f')^2}{2}
+ f^2 n^2 \frac{N^2-L^2 W^2}{2 N^2 L^2}\ \ , \nonumber \\
\ \ T^t_{\phantom{0}\phi} &=&  \frac{n f^2 (\omega  + n W)}{N^2}
\end{eqnarray}
\subsection{Ricci tensor}
Correspondingly, the relevant components of the Ricci tensor reduce to
\begin{eqnarray}
R_t^t &=& \frac{(KLN')'}{KLN} + W W' \frac{L(KLN' - 3 K N L' - K' N L)}{2 K N^3}  -((W')^2 + W W'') \frac{L^2}{2 N^2}\ \ , \ \ \nonumber \\
R_r^r &=&  \frac{N''}{N} + \frac{K''}{K} + \frac{L''}{N} - (W')^2 \frac{L^2}{2 N^2} \ \ , \ \ \nonumber \\
R_{\varphi}^{\varphi} &=&   \frac{(KL'N)'}{KLN} - W W' \frac{L(KLN' - 3 K N L' - K' N L)}{2 K N^3}+ (W'' W + (W')^2) \frac{L^2}{2 N^2}   \ \ , \ \ \nonumber \\
R_z^z &=& \frac{(LNK')'}{KLN}\ \ , \ \ \nonumber \\
R^t_{\phantom{0}\phi}&=& \frac{L^2 W''}{2 N^2} - \frac{L W' ( K L N' - 3 K L' N - K' L N)}{2 K N^3} \ \ .
\end{eqnarray}

\subsection{Physical quantities}
The solutions can be characterized  by their energy $E$ and tension $T$ per unit lenght.
Along with \cite{p_peter,Hartmann:2012mh}
we adopt for these quantities the definitions
\begin{equation}
E = \int \int \sqrt{-h} \ T_t^t dR d \varphi \ \ \ , \ \ \ T = \int \int \sqrt{-h} \ T_z^z dR d \varphi \ ,
\end{equation}
where $h$ represents  the determinant of the induced metric on the $R,\varphi$-plane (the integration over the angle $\varphi$ is trivial). Note that another definition of the mass is used in \cite{Verbin:1998tc,Christensen:1999wb}. 
Solitons like Q-balls and Q-tube can further be characterized by the Noether charge underlying the global symmetry
of the model
\be
       Q = \int j^0 \sqrt{-g} d\varphi dR = 4 \pi \int_0^\infty f^2
\frac{\omega+nW}{N^2} NLK dR \ \ ,
        \ \ 
\ee
In the case of spinning soliton the solution can further be characterized
by the  angular momentum $J$ (per unit lenght). The quantities $J$ and $Q$ are proportional~:
\be
           J = \int T^0_2 \sqrt{-g} dR d\varphi = 4 \pi n \int_0^{\infty} N L K [f^2 \frac{\omega + n W}{N^2}] dR
           \ \ ,  \ \ J = n Q
\ee
For $Q$ and $J$, we used the  definition of the measure which makes them proportional. 
\section{Solutions for $n=0$}
\begin{figure}[h!]
\begin{center}
\subfigure[Energy, tension, charge]
{\label{kappa04_l}\includegraphics[width=8cm]{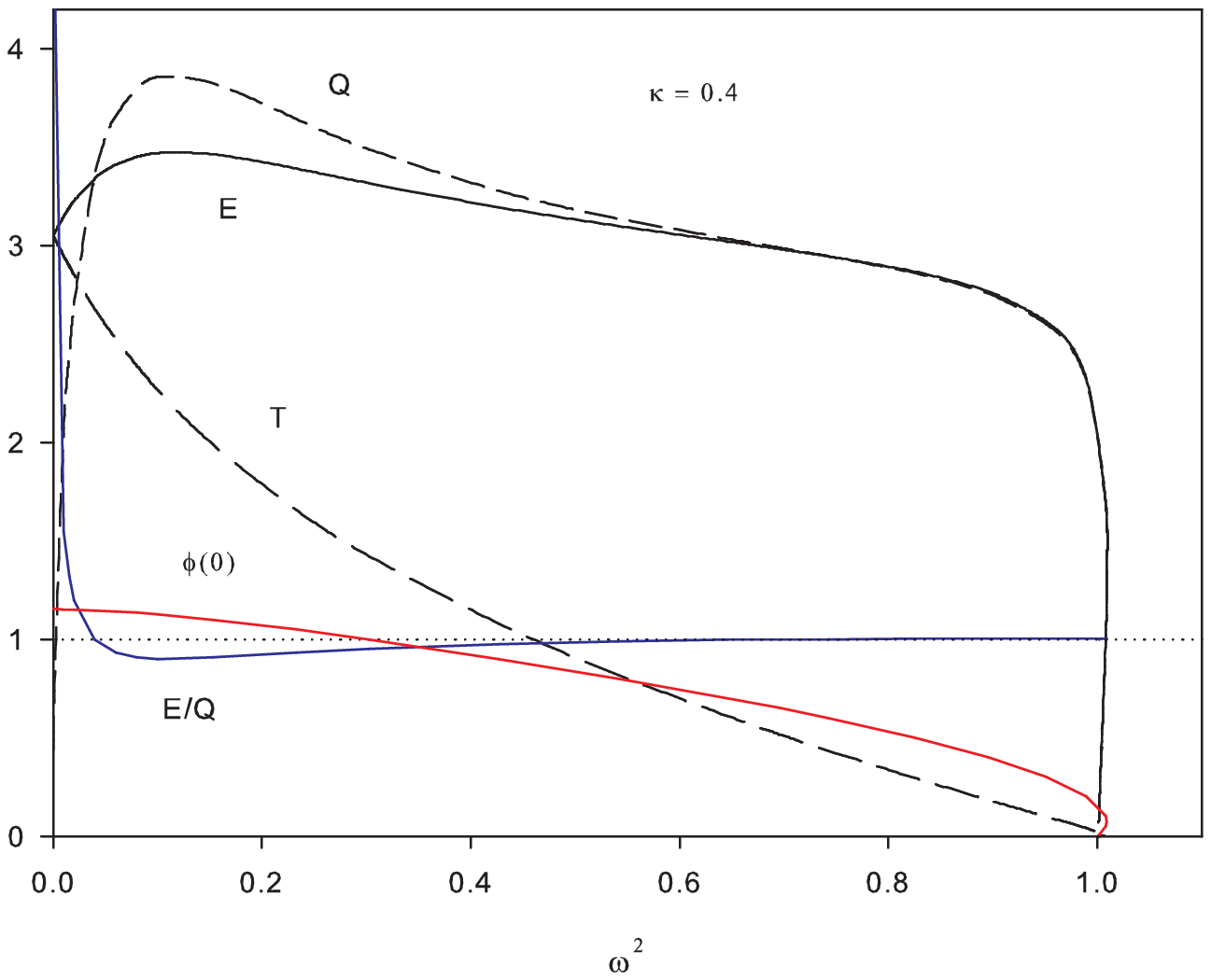}}
\subfigure[Kasner parameter and angular deficit]
{\label{kappa04_r}\includegraphics[width=8cm]{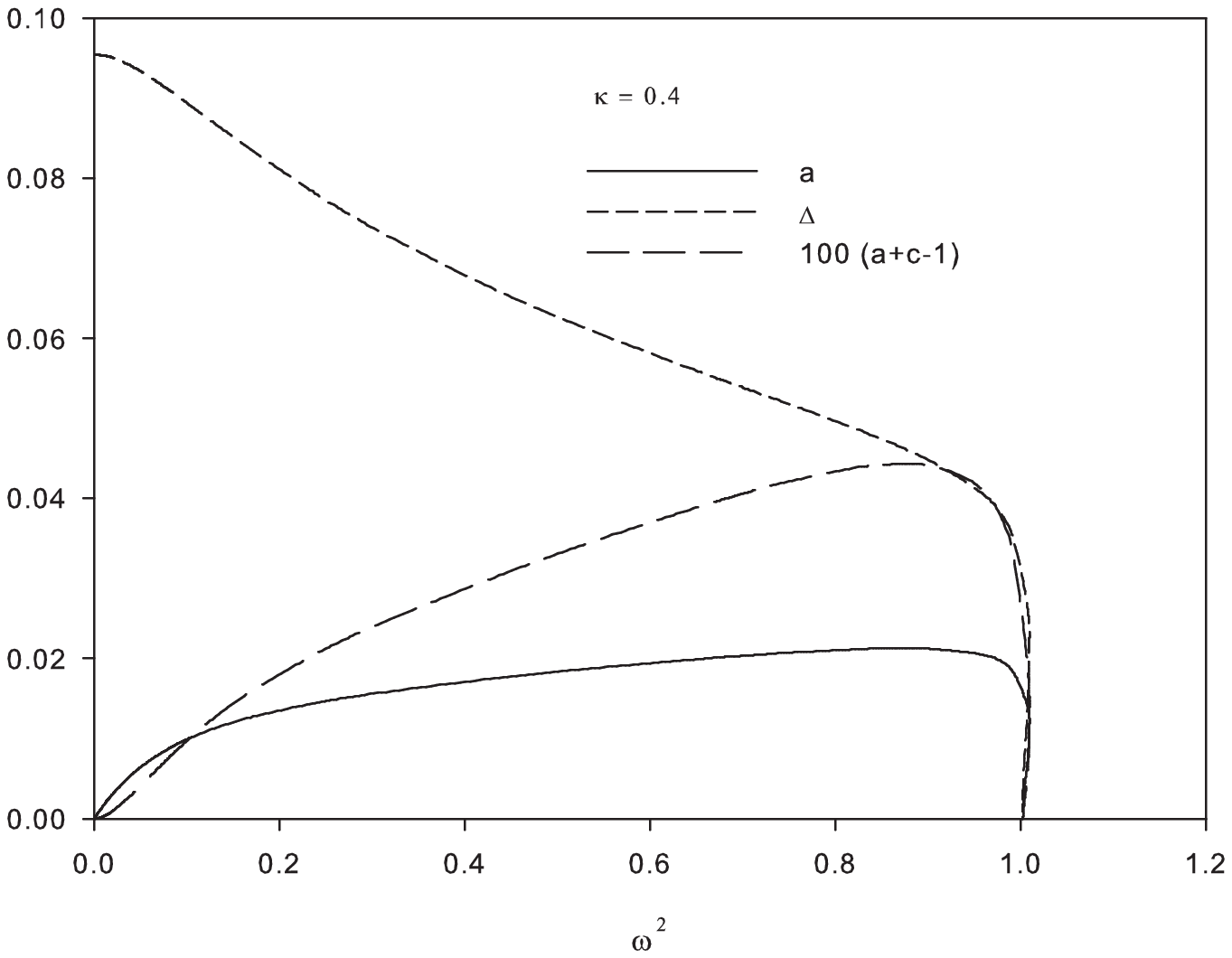}}
\end{center}
\caption{The quantities $E$, $T$, $Q$ (left);  the Kasner power and angular deficit $\Delta$ 
as functions of $\omega^2$ for $n=0$, $\alpha=0.1$ and $\kappa = 0.4$. 
\label{fig1}
}
\end{figure}
To our knowledge, the coupled field equations do not admit solutions in an explicit form.
We relied on numerical techniques to
construct them. The routine Colsys \cite{colsys} was used for this purpose.
By an appropriate rescaling of the scalar field, the parameter $\gamma$ of the potential in (\ref{lmatter}) can be set
to one and the mass parameter $m$ can be absorbed in the gravitation constant, 
we therefore conveniently use $\alpha = 16 \pi m^2 G$. With these scale conventions, the system appears with
two intrisically different coupling constants : $\alpha$ and $\kappa$. 
Families of solutions labelled by $\omega$ and $\lambda$ 
(or alternatively by $f(0)$) can then be constructed numerically.  
We first concentrate on the-non spinning case characterized by $n=0$. 
Assuming for the moment that there in no boost (i.e. setting $\lambda=0$),
the conditions associated to the non diagonal Einstein equations are solved consistently by setting $W=v=M=0$.
The boundary conditions which guarantee the regularity of the configuration on the symmetry axis  (i.e. for $R=0$)
and the finiteness of the energy (per unit length with respect to $z$) are
\be
\label{boundary_conditions}
  N(0) = 1 \ \ , \ \ N'(0) = 0 \ \ , \ \  K(0) = 1 \ \ , \ \ K'(0) = 0 \ \ ,  \ \ L(0) = 0 \ \ , \ \ L'(0) = 1 \ \ , \ \ 
  f'(0) = 0  \ \ , 
\ee
while $f(0)\equiv \psi_0$ is a free parameter 
(note: the conditions $N(0)=K(0)=1$ result of an appropriate rescaling of the coordinates $t$ and $z$). 
The asymptotic condition $f(R \to \infty) = 0$ completes the boundary value problem. 
In the absence of gravity, the scalar field equation
implies the following exponential decay of the scalar field
\begin{equation}
 f(R) \propto   \frac{1}{\sqrt R} e^{- \sqrt{m^2-\omega^2}R} \ \ .
\end{equation}
For $\alpha > 0$, the asymptotic form is more involved but the numerical result confirm that
the scalar field indeed  decays exponentially.
\begin{figure}[h!]
\begin{center}
\subfigure[$N$ and $K$]
{\label{profile_nk}\includegraphics[width=8cm]{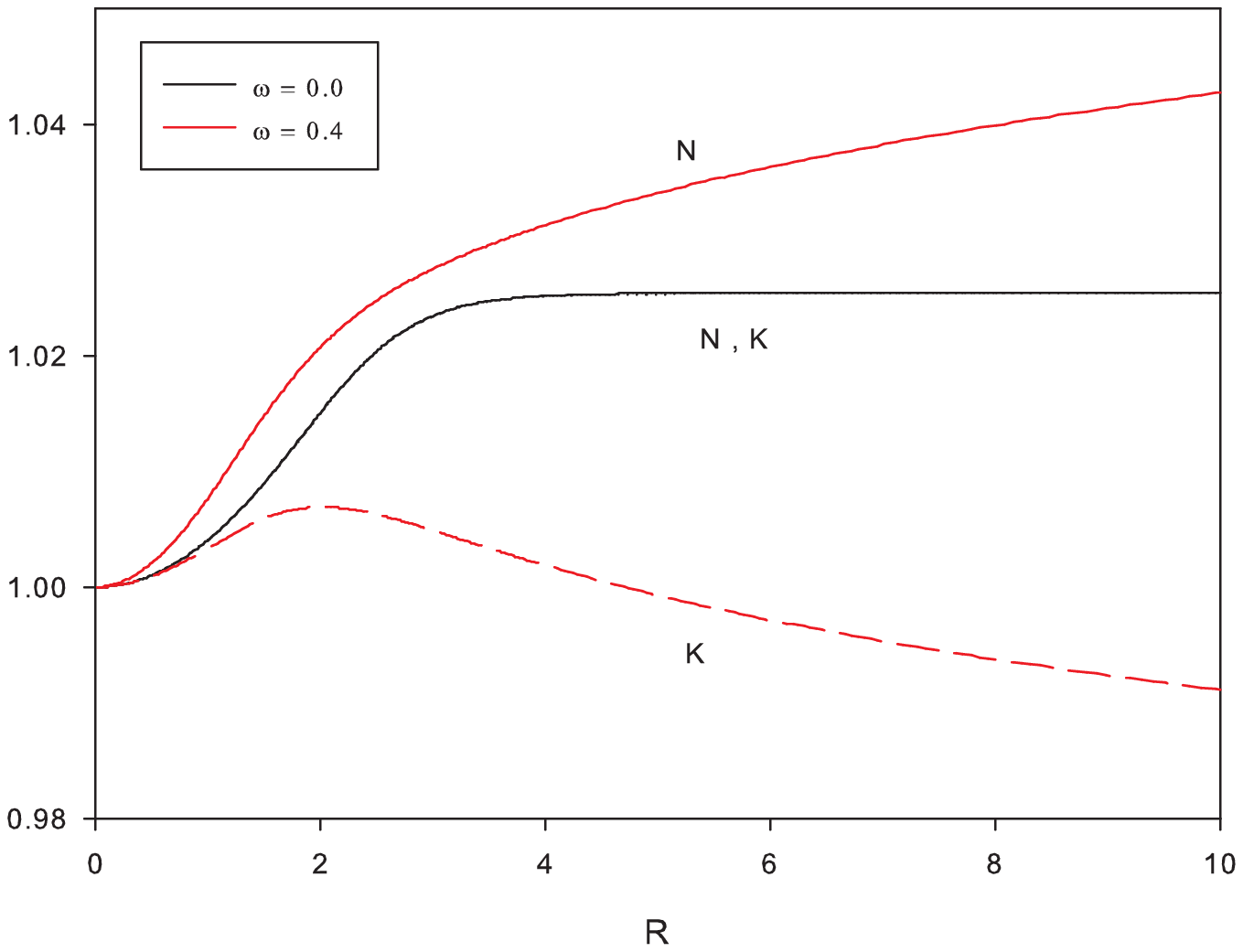}}
\subfigure[$L'$ and $\phi$]
{\label{profile_phil}\includegraphics[width=8cm]{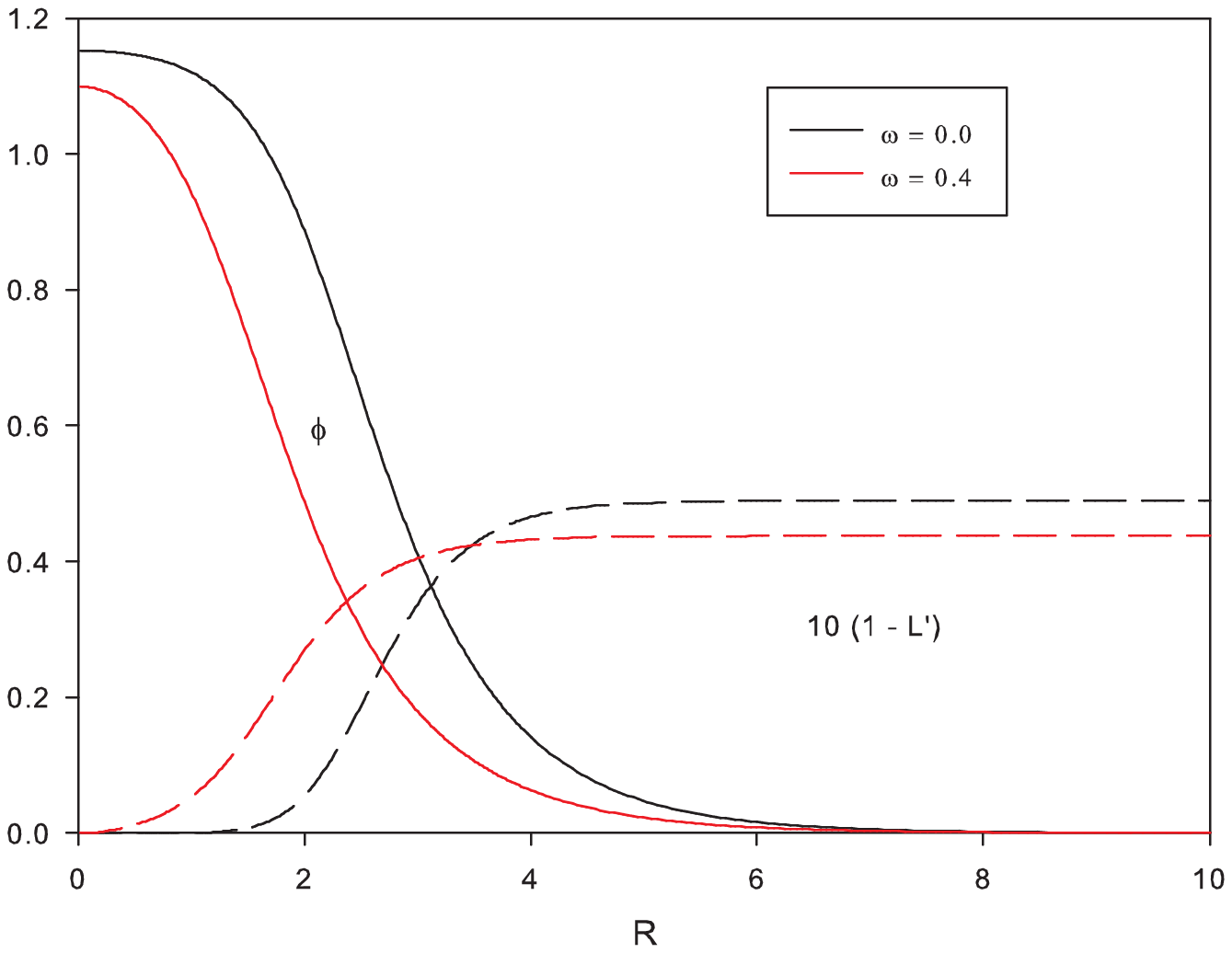}}
\end{center}
\caption{Comparaison of profiles of the metric functions $N,K$ (left),  
$L'$ and the scalar field (right) for  $\omega=0.0$ and $\omega=0.4$ (here
$\alpha=0.1$, $\kappa=0.4$).
\label{fig2}
}
\end{figure}

Fixing a value for $\omega$ in principle leads to one (or more, see below) solution with a fixed value $\psi_0$. 
We found it convenient to supplement the system with the equation $d \omega / dR = 0$ 
and to take advantage of the extra boundary condition to impose $f(0) = \phi_0$. 
The corresponding value of $\omega$ being  reconstructed numerically.
With the potential specified as above, the reasonning of Ref. \cite{Volkov:2002aj} 
demonstrates that the solutions exist in the interval
\begin{equation} 
              \omega_c^2 \leq \omega^2 \leq m^2  \ \ , \ \  
              \omega_c^2 = {\rm min} \{ 0, \frac{m^2}{2}(1 - \frac{1}{2 \kappa}) \}
\end{equation}
The pattern of solutions is therefore 
 different for $\kappa < 1/2$ and $\kappa > 1/2$. 
\begin{figure}[h!]
\begin{center}
\subfigure[Mass, tension, charge]
{\label{kappa04_l}\includegraphics[width=8cm]{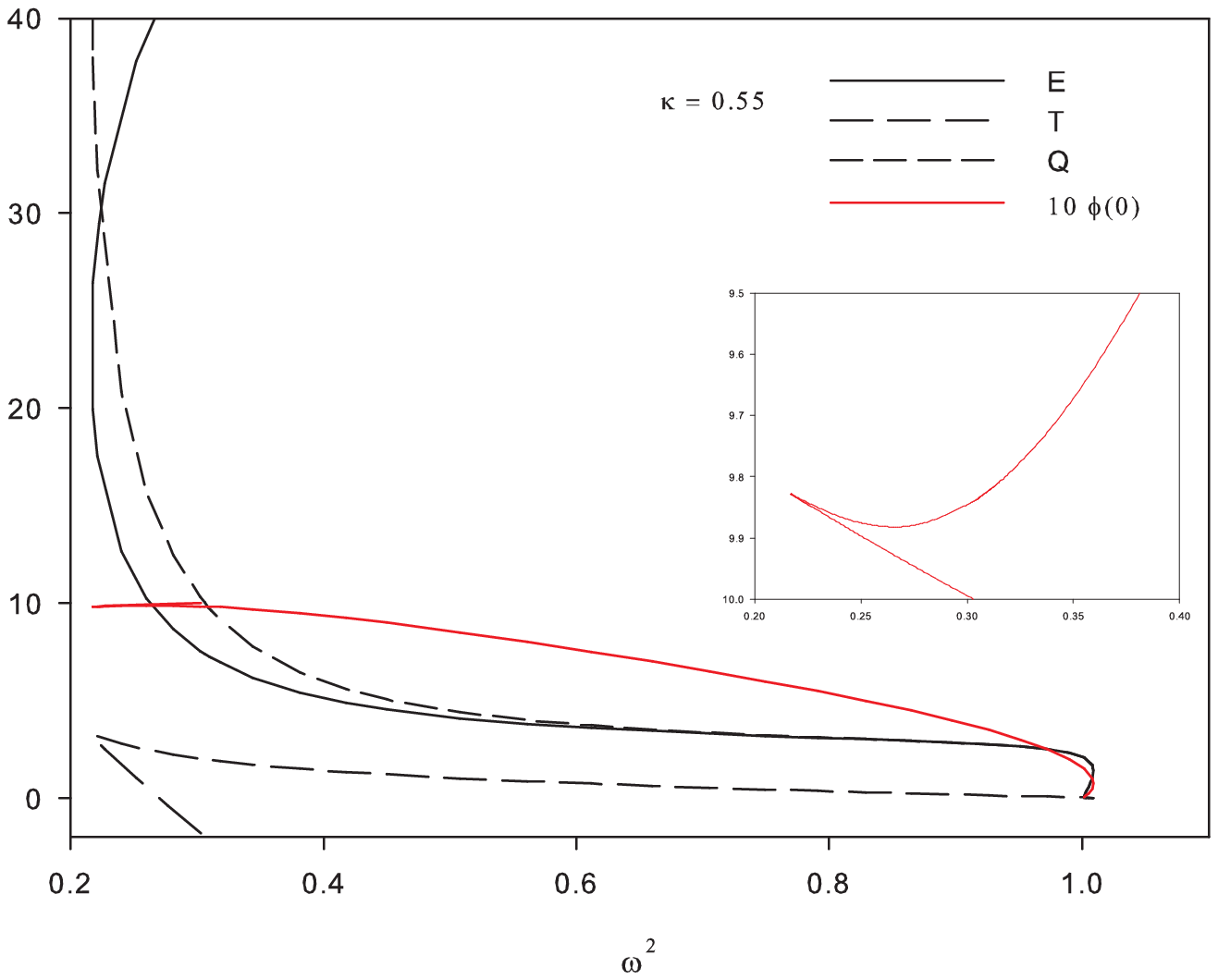}}
\subfigure[Kasner parameter and angular deficit]
{\label{kappa04_r}\includegraphics[width=8cm]{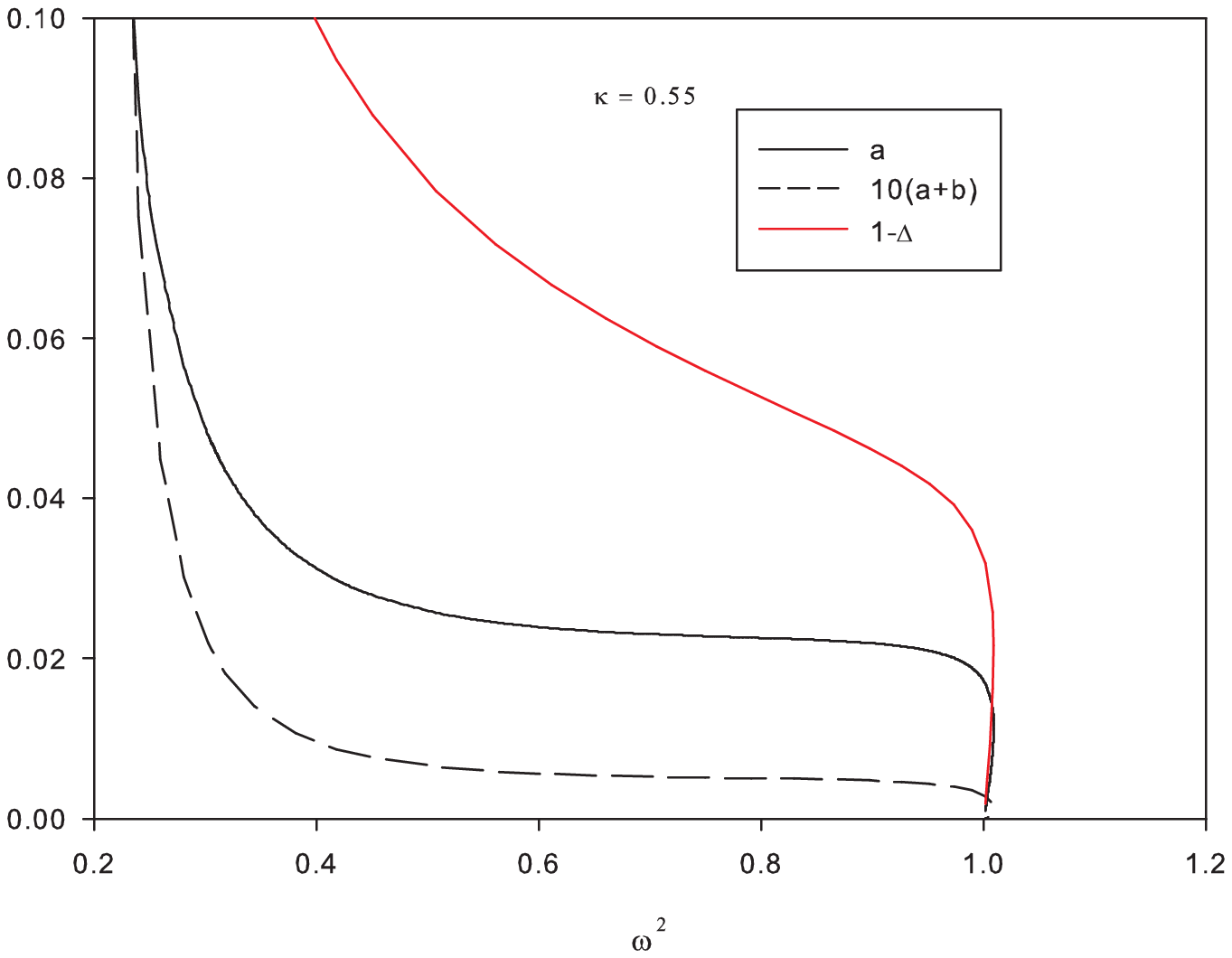}}
\end{center}
\caption{The quantities $E$, $T$, $Q$(left); the Kasner power and angular deficit $\Delta$ 
as functions of $\omega^2$ for $n=0$, $\alpha=0.1$ and $\kappa = 0.55$. 
\label{fig3}
}
\end{figure}

\subsection{ $\kappa < 0.5$}
In this case the soliton can be constructed up to $\omega = 0$. Setting for definiteness $\kappa = 0.4$ and $\alpha=0.1$, 
we obtained the family of solutions whose pattern is illustrated by Fig. \ref{fig1}.
We see that, for $\omega^2 \rightarrow 1$, the scalar field approaches the vacuum (i.e. $\phi(x) = 0)$ and  Minkowski
space-time is approached.
For $\omega \to 0 $, the matter field in the core becomes important and the angular deficit is large. 
For the value $\omega = 0$ the metric functions $N(R),K(R)$ become constants for $R \to \infty$ while the angular deficit is maximal.  
The profiles of the solutions are shown on Fig. \ref{fig2} and compared for two different values of $\omega$.
\subsection{ $\kappa < 0.5$}
The crucial difference with respect to the case $\kappa > 0.5$ is that the solutions stop at a strictly positive value of $\omega$, say $\omega=\omega_c$.
In fact it turns out  that for small enough values of $\omega^2$ two solutions exist forming two branches that coincide at $\omega=\omega_c$. 
It is natural to refer to the branch directly connect to Minkowski space-time as to the main branch since, for a fixed $\omega$, it has the lowest mass.
When we progress on the second branch (while increasing $\omega$) the so called 'thin wall limit' is approached. In particular, the
scalar function varies very slowly for $R < R_c$ for some definite radius $R_c$ and then rapidly reaches 
its asymptotic value $\phi = 0$ at $R \sim R_c$.
At the same time, the charge and energy (per unit lenght) become quite large while the tension becomes negative.
These properties are illustrated on Fig. \ref{fig3}.
\subsection{Boosted solutions}
All the solutions available in the case $n=0$ can be deformed by a boost, i.e. setting $\lambda > 0$ in the equations.
The Einstein equation corresponding to $R_t^z$ is no longer trivial leading to a non vanishing function $v(R)$.
It turns out that the function $v(R)$ and the parameter $\omega$ 
enter the equations through a special combination which we conveniently redefine as
$\tilde v(R) \equiv v(R)+\omega/\lambda$.
 The relevant boundary to solve
the system of five equations consists of the  conditions
(\ref{boundary_conditions}) for the metric functions $N,K,L$ supplemented by
$$
      f(0) = \phi_0 \ \ , \ \ f'(0) = 0 \ \ , \ \ \tilde v'(0) = 0 \ \ , \ \  f(R \to \infty) =0 \ \ .
$$
where $\phi_0$ is a constant. This leads to a family of solutions labelled by the two parameters $\lambda$ and $\phi_0$.
These calculations are  systematic and not reported here.
\section{Solutions for $n>0$}
Setting $n>0$, the above equations lead to spinning solutions since the component $T^0_{\varphi}$
of the energy momentum tensor is non zero. 
The regularity of the scalar field equation at the origin imposes $f(0)=0$ while
the derivative $f'(0)$ is determined by the numerical result.
\subsection{Spinning Q-tubes}
Setting $\alpha = 0$, we have constructed  families spinning solutions -without gravitation- of the field equations.
The two dimensional counterpart was obtained in \cite{Volkov:2002aj}.
In this case, the Einstein equations  lead to a Minkowski space-time (i.e. with $W=0$).
Along with the corresponding solutions in the $n=0$ case, the spinning Q-tubes exist for a finite interval of the parameter $\omega$.
The physical parameters of the solutions  corresponding $\kappa = 0.4$ and $\kappa = 0.55$ are reported on Fig. \ref{fig4}.
When the solutions cannot be constructed in the limit $\omega = 0$, they exist up to a minimal value of $\omega$
(for $\kappa = 0.55$, we find $\omega_{m} \approx 0.38$). When the minimal value of $\omega$ is approached the solution
approaches a thin wall limit configuration, both the energy and the charge become quite large as illustrated by Fig. \ref{fig4}.
As pointed out above, the relation $J = Q$ holds for these non gravitating Q-tubes. 
\begin{figure}[h!]
\begin{center}
\subfigure[Mass, tension, charge $\kappa=0.4$]
{\label{n_1_kappa04}\includegraphics[width=8cm]{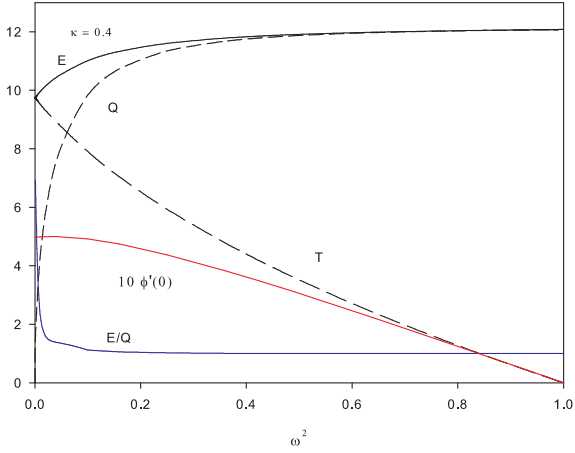}}
\subfigure[Mass, tension, charge $\kappa=0.55$]
{\label{kappa04_r}\includegraphics[width=8cm]{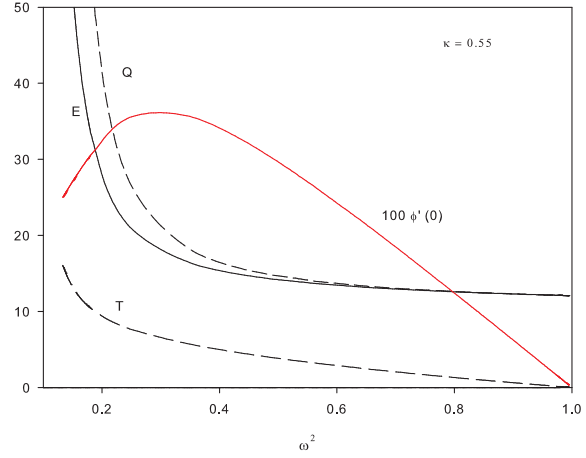}}
\end{center}
\caption{The parameters $E$, $T$, $Q$ per unit lenght  
as functions of $\omega^2$ for $n=1$, $\alpha=0$ and $\kappa = 0.40$ (left),  $\kappa=0.55$ (right). 
\label{fig4}
}
\end{figure}
\begin{figure}[h!]
\begin{center}
\subfigure[Profiles $\kappa=0.55$]
{\label{n_1_kappa04}\includegraphics[width=8cm]{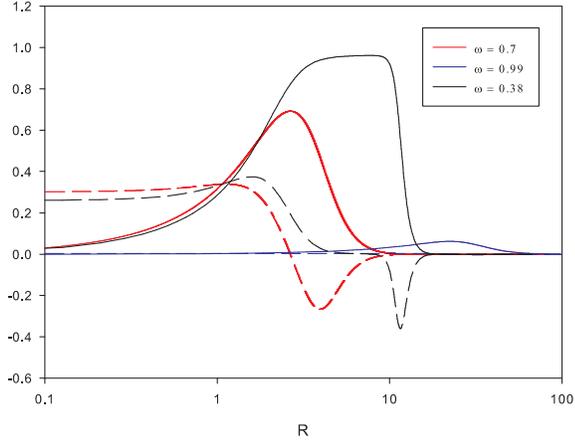}}
\subfigure[Profiles $\kappa=0.4$]
{\label{kappa04_r}\includegraphics[width=8cm]{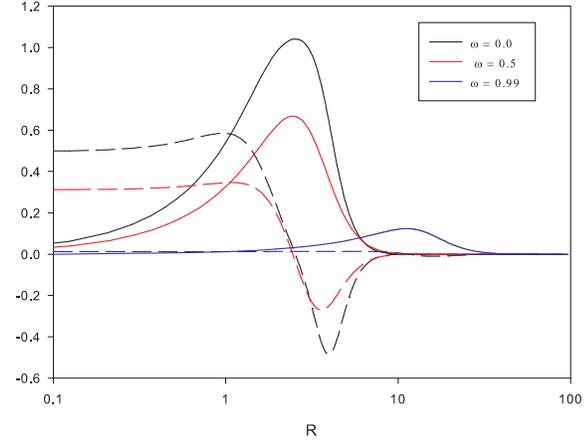}}
\end{center}
\caption{Profiles of $f,f'$ (solid and dashed lines respectively) for spinning Q-tubes for three values of $\omega$. 
\label{fig_N_1}
}
\end{figure}
\subsection{Gravitating, Spinning Q-tubes}
Setting $n=1$ and $\alpha > 0$ in the equations does not lead to any reduction of the system
for generic value of $\omega$. Only in the case $\omega = 0$ the equation for the function $W(R)$ 
becomes linear and is fullfilled by $W(R)=0$. 
For the metric fields $N,L,K$, the boundary conditions on the symmetry axis are choosen like 
in the non spinning case.
The regularity of the equations at $R=0$  further requires $W'(0)=0$;
it is  natural to complete the boundary boundary value problem by imposing $W(R \to \infty)=0$. 
The numerical results confirm the existence of regular solutions extrapolating between the initial conditions
imposed at $R=0$ and the asymptotic Kasner form. 

As pointed out in Sect. 2.2, the scalar field is exponentially small  in the asymptotic region.
  In this region a metric of the form (\ref{gparticular}) in particular $W$ obeys Eq. (\ref{asymptotic_w}).
where $C$ and $W_0$ can be determined numerically.  
The three powers $a,b,c$ can be extracted from the numerical solutions and it was confirmed
that the obey the Kasner relations within the numerical errors (of order  $10^{-6}$ in our case). 
With the value $\alpha = 0.1$ used to compute the solutions, the Kasner powers are such that the terms depending on
$W(R)$ in the Einstein equations  are  subleading  and the effect of the rotation can be considered
as a  correction with respect to a purely Kasner solution.
In particular, it turns out  that the Kasner powers  are such that the relation $2b+c>1$ is obeyed 
 for the families of solutions that we have studied.
 (more precisely $0 \leq a \leq  0.08$; $ 0.9925 \leq b \leq 1$ ; $0 \geq c \geq -0.082$, 
 they are reported on Fig.\ref{spinning_0}). In spite of the fact that the diagonal terms are dominant in the equations,
  the non-diagonal component of the metric $g_{02}$ does not vanish in the asymtotic region.
The spinning Q-ball somehow produces a rotationnal effect on space-time which  perists in the asymptotic region.

The fact that the  scalar field approaches the asymptotic value $f(R\to \infty)=0$  exponentially ensures
that the integral determining the quantities $U,T,Q,J$ are convergent. 
The profiles of $T^0_0$, $T^0_2$, $g_{00}$, $f(R)$ and $W(R)$ are shown on Fig. \ref{spinning_0} (left side)
 in the case
$\alpha = 0.1$, $\omega = 0.5$. The dependance of the mass, the tension and the angular momentum on the parameter
$\omega$ are reported on the right side of the figure.
\begin{figure}[h!]
\begin{center}
\subfigure[Energy momentum, metric and scalar field]
{\label{spinning_0_a}\includegraphics[width=8cm]{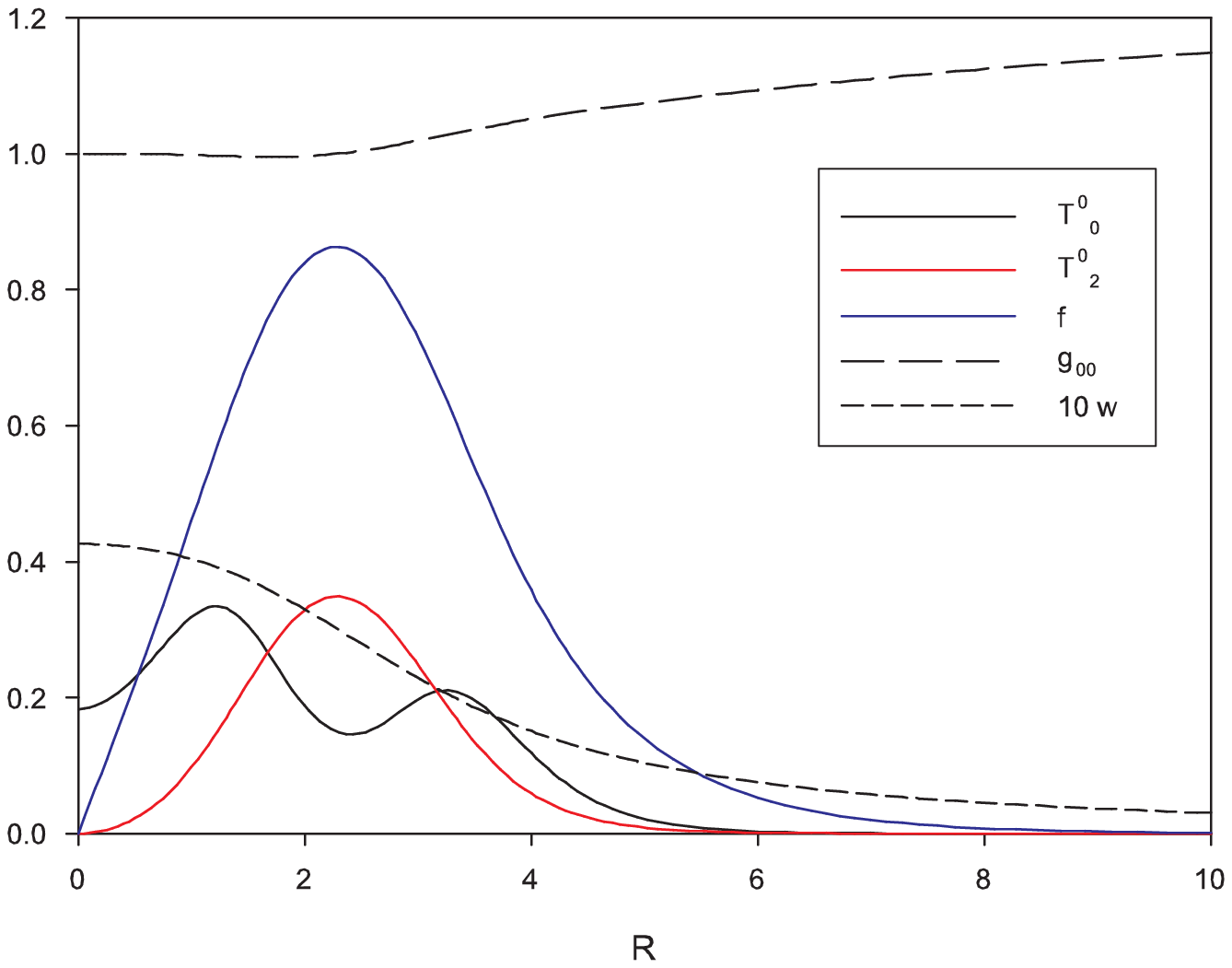}}
\subfigure[Mass, tension and angular momentum]
{\label{spinning_0_b}\includegraphics[width=8cm]{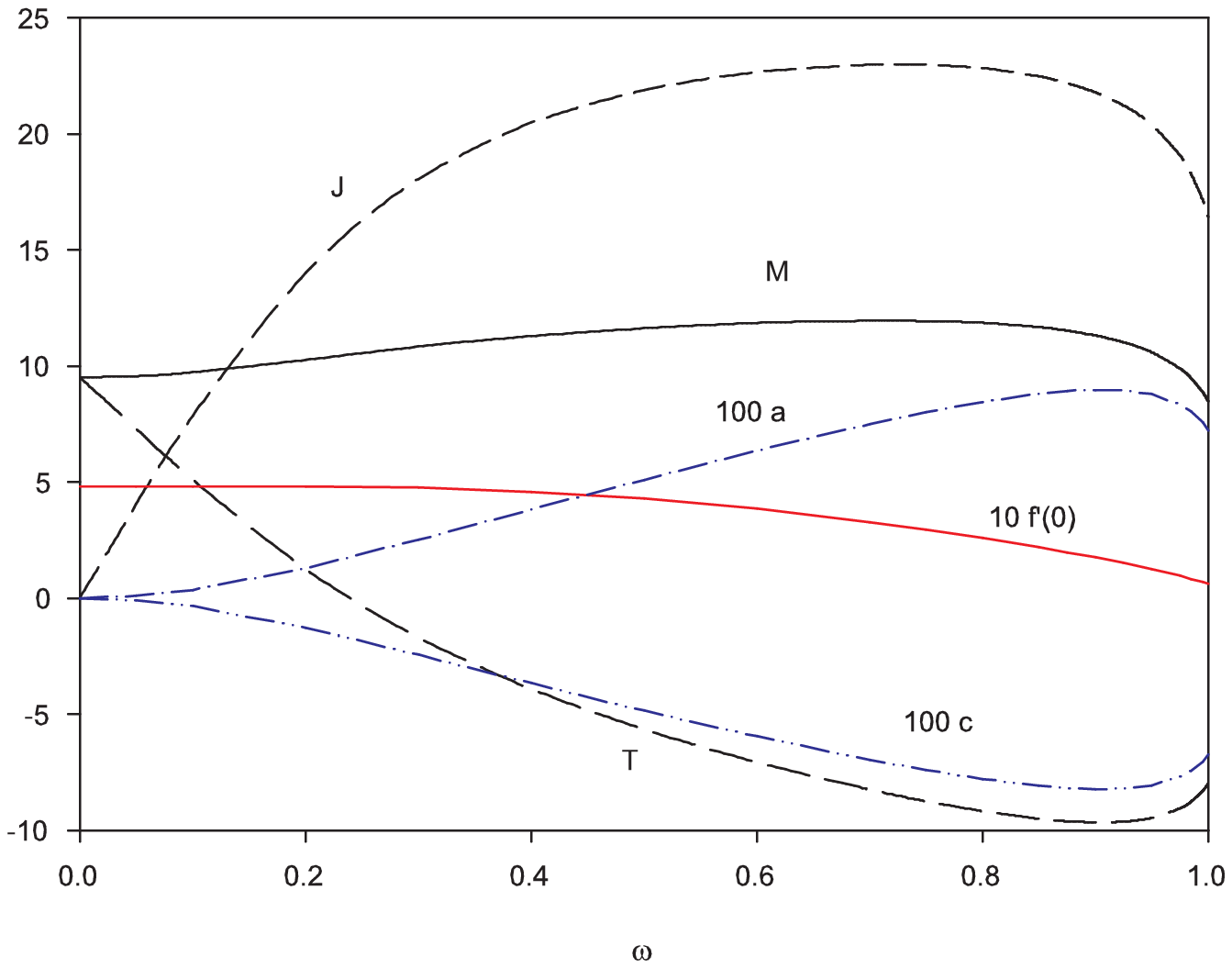}}
\end{center}
\caption{The metric components $g_{00}$, W(R), f(R), the energy and angular momentum
density for spinning, gravitating Q-tube ($n=1$, $\alpha = 0.1$) for $\kappa=0.4$ and $\omega=0.5$ (left). 
Mass, Tension, angular momentum and the parameters $a,c$ as functions of $\omega$ (right). 
\label{spinning_0}
}
\end{figure}

Independently of the solutions obeying the boundary condition $W(R\to \infty)=0$, 
we produced another family  of spinning Q-tubes by imposing for the metric function $W(R)$ the condition $W(0)= W'(0)=0$.
In this case also, the scalar field is localized and the asymptotic value $f=0$ 
is appoached exponentially. Outside the core of the string,
the different metric fields behave according to
\begin{equation}
\label{asymptotic}
N(R) = N_0 R^a \ \ , \ \ L(R) = \Delta R \ \ , \ \ K(R) = K_0 R^{-a} \ \ , \ \ W(R) = W_0 + o(R^{2a-2})
\end{equation}
where the parameters $N_0,K_0,W_0$, $\Delta$ and the power $a$ are constants. 
These results reveal that the time-time component of the metric $g_{00} = N_0 R^{2a}-W_0^2 \Delta^2 R^2$
vanishes for a critical value of the radius $R= R_c$, where $R_c$ depends on the parameters.

Some relevant metric components of the  solution of this type corresponding to $\alpha = 0.1$, $\omega = 0.05$ 
are presented on Fig. \ref{spinning_1}; in this case, we have $R_c \approx 176$,
$a\approx 0.0045$, $\Delta \approx 0.848$. 
For the domain of the parameters that we have explored, the scalar field reaches its asymptotic value
already for $R \ll R_c$; the form (\ref{asymptotic}) therefore corresponds to a vacuum solution.
However, it does not belong to the subclass (\ref{gparticular}).
\begin{figure}[h!]
\begin{center}
\subfigure[Ricci, Kreischman scalar and $g_{00}$]
{\label{spinning_1_l}\includegraphics[width=8cm]{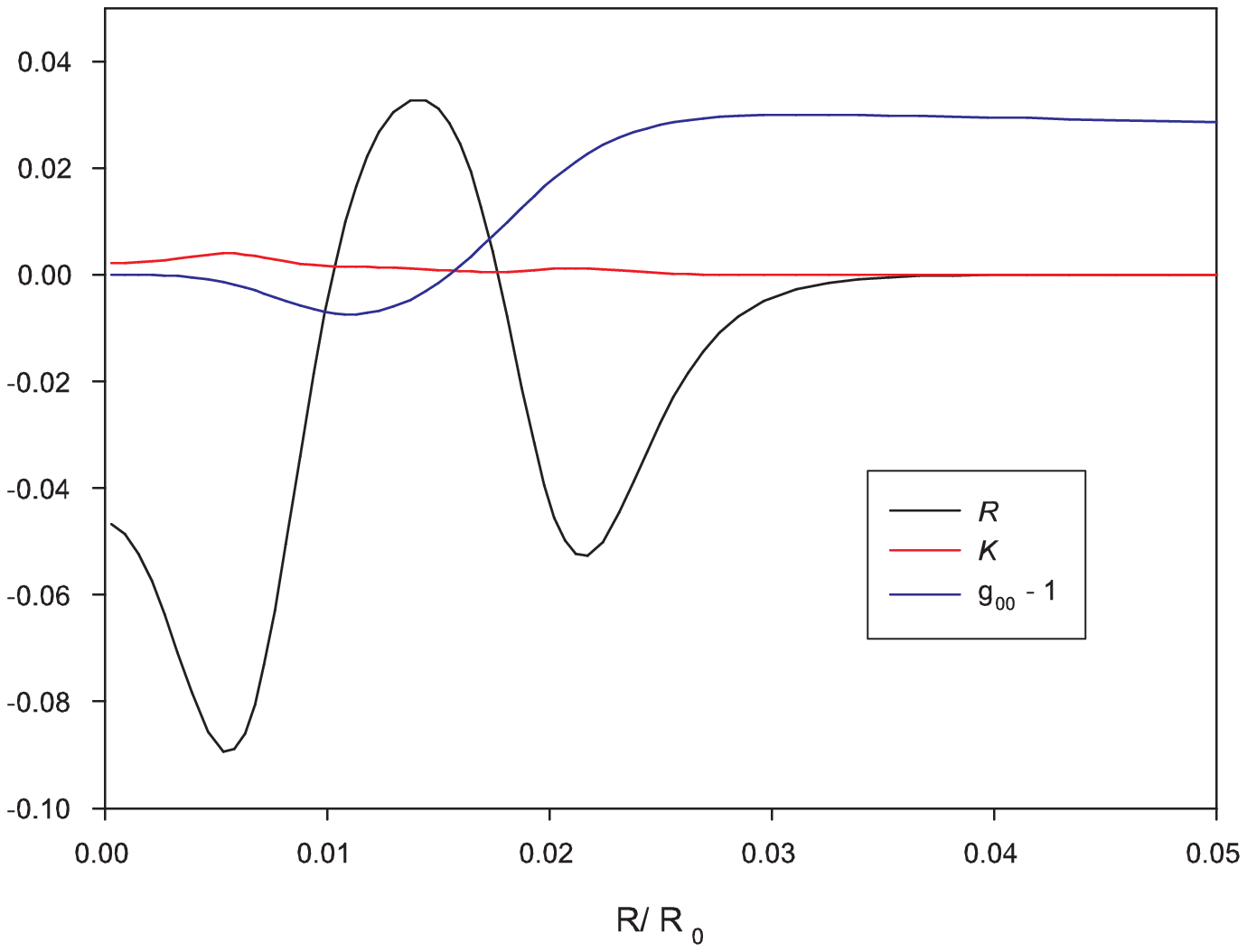}}
\subfigure[Metric components]
{\label{spinning_1_r}\includegraphics[width=8cm]{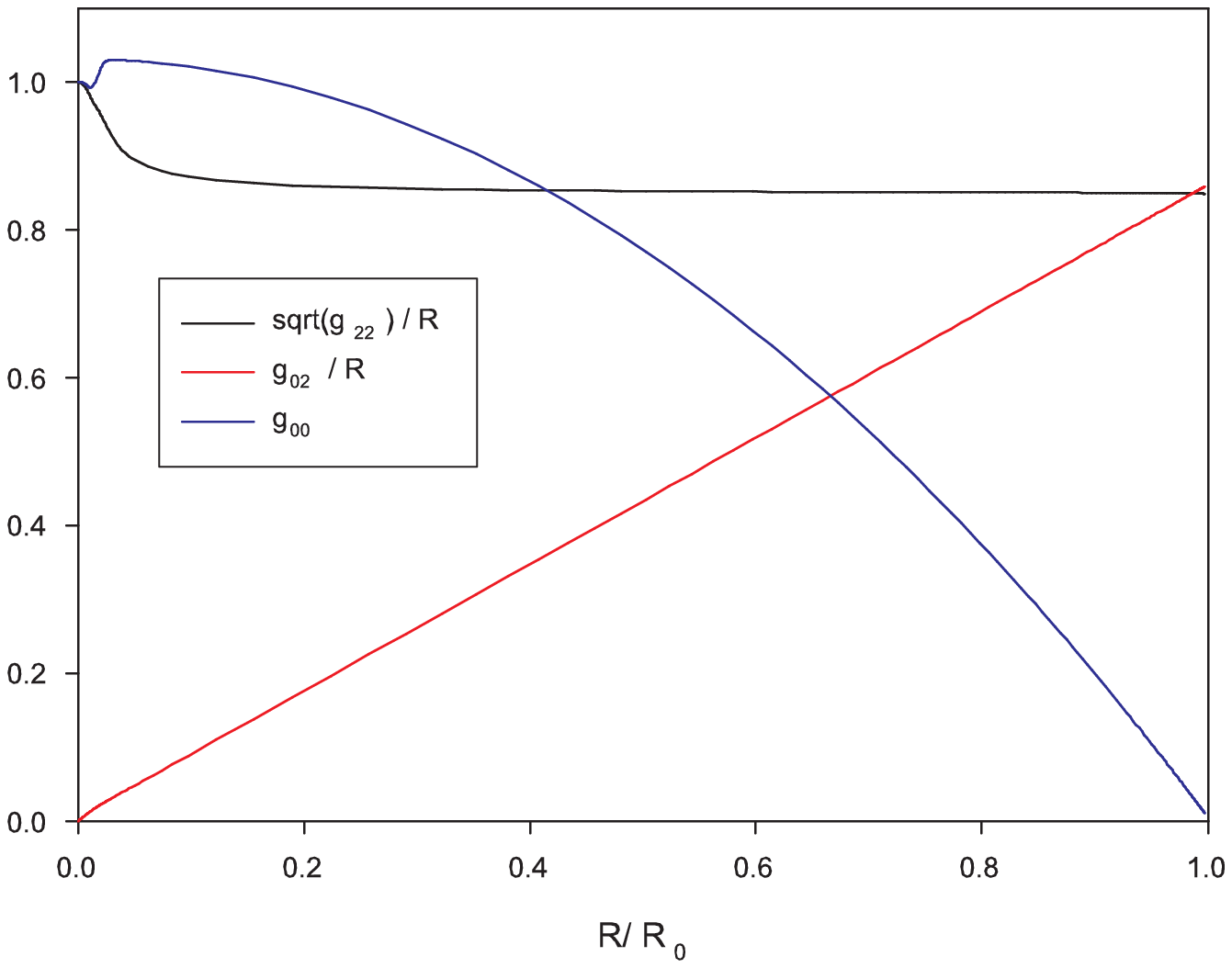}}
\end{center}
\caption{Profiles of the metric components, of the Ricci and Kreischmann invariants
 of a spinning gravitating Q-tube ($n=1$,$\alpha = 0.1$) for $\kappa = 0.4$ and $\omega = 0.05$.  
\label{spinning_1}
}
\end{figure}
At the approach of the limiting point $R=R_c$  the  metric takes the form
\begin{equation} 
   ds^2 = (R_c-R) dt^2 - \Delta^2 R^2 d\varphi^2 -  dR^2 - K_0^2 dz^2 - W_0^2 \Delta R dt d \varphi \ \ , 
   \ \  
\end{equation}
The  solutions under consideration  are then regular only for  $R < R_c$. 
The apparent singularity occuring for $R \to R_c$ 
is in fact a coordinate artefact, the Ricci and Kreischmann 
scalar invariants indeed vanishes in the limit $R\to R_c$ as demonstrated  on Fig. \ref{spinning_1}.
It seems that, far away from the center of the lump,
the time of space-time becomes degenerate due to the spinning nature of the Q-tube ($W_0 > 0$, $\omega > 0$).  
These spinning solutions share the property that $g_{00}(R_c)=0$ with the 
'singular Kasner' solutions of the classification reported in \cite{Christensen:1999wb}. 
However in the case of the singular Kasner the metric component $g_{22}(R)$ gets singular for $R \to R_c$
while it remain regular  (for instance $g_{22}(R) \sim R^2$) in the present case.   

The underlying space-time is also different from the space-time of supermassive cosmic strings 
discussed e.g. \cite{laguna,ortiz}. The associated metric function $g_{22}(R)$ 
indeed gets a zero at some radial radius, corresponding
to a maximal angular deficit.

\section{Conclusion}
In this paper, we constructed numerically the gravitating counterpart of
the Q-tubes of \cite{Sakai:2010ny} including  rotation or boost. 
This was done by using an appropriate ansatz for the fields.
In particular, the metric degrees of freedom are encoded in six independant
functions of a coordinate $R$ representing the distance to the axis of symmetry. 

As a preliminary setup for the understanding of the solutions,
we  analyzed the structure of the vacuum Einstein equations underlying
a boosted-rotating tube. 
The construction elaborated  in \cite{clement} leads to analytical solutions
that can be adapted to our parametrisation of the metric.
We manage to construct -within our ansatz- the  six first integrals to the system transforming the original
system of six {\it second} order equations  into a system of six  {\it first} order equations. 
Two of these 'constants of motion' can be interpreted as the angular momentum and the linear
momentum along the axis of symmetry (in our notations $K_\varphi$ and $K_z$ respectively).
The interpretation for the
other constants is less clear, although the energy of the configuration
must contain the sum $K_t +K_{zz}  + K_{\varphi\varphi}$.

Our results demonstrate that the $Q$-tubes constructed in flat space in
\cite{Sakai:2010ny} are smoothly deformed by gravity, leading to a space-time
that is asymptotically of Kasner type.
The matter field corresponding to the Q-tube is localized around the axis
of symmetry and quickly approaches its asymptotic value.
The solitons with $n=1$ are spinning 
in the plane transverse to the axis of symmetry, while the boost corresponds to
a flux along the direction of the axis. 

It turns out that for the solely rotating or solely boosted solitons the
asymptotic metric is a Kasner spacetime  corrected by subdominant terms. 
We obtained several types
of spinning, gravitating Q-tubes, labelled essentially by  $W(0)$, i.e.
the  value on the axis of the rotating function on the axis.
For a subset of the solutions obtained,  the time-time component of the metric vanishes 
at a critical distance, say $R=R_c$, to the axis of symmetry; the value $R_c$ is 
much higher than the soliton core. We believe that this feature is related to the fact that the
string is infinite and that it would be regularized in the case of stings of finite size.

We obtained many examples of regular solitons with  both rotation and
boost, although they seem to be regular only on a small range of the parameters. 
When the boost parameter $\lambda$ becomes too large,  some components of the metric change
sign (or become singular) at  finite values of the radius. 
In this case, the metric seems not to be of  Kasner type {\it even} up to subdominant terms.
Our results shows that, in all cases, the diagonal components of the metric obey
 one of the two Kasner conditions (the linear one), while the quadratic relation is not fullfilled. 
The full understanding of the pattern of solutions needs further
investigations which we delay to future considerations. 

Finally, note that the metric \eqref{metric} could be suitable to describe the
near zone of a solitons with a torus shape (like vorton, see \cite{radu_volkov}), 
similar to the black ring  constructed in \cite{Emparan:2007wm} in pure gravity.
In this case, the  black string with a momentum along the axis describes the near zone of
the ring. Intuitively, the momentum along the axis is mapped to the angular
momentum of the ring and balances the self gravity tending to collapse the
ring. We could expect a similar phenomena here since quantities like tension and
angular momentum are naturally defined from the matter fields. 

For cosmic string solutions criteria have been developed to decide 
about the stability of these objects which is crucial for the
possible formation of vortons \cite{carter_paper}. 
This uses the macroscopic properties of the strings in the sense that
the velocities of the longitudinal and transversal perturbations 
are determined in terms of the energy per unit length and the tension
of the string. This method has been used to decide about 
the stability of superconducting cosmic strings 
(see e.g. \cite{p_peter}, \cite{Hartmann:2012mh}, \cite{babeanu_hartmann})        
It would be interesting to see whether these techniques can also 
be employed in our case since the criteria used only rely on the macroscopic
quantities of the string.
\appendix 
\section{Stress tensor in the general case}
\label{app:eqs}
The Einstein tensor in the general case is easily reconstructed from
\eqref{firstint}. We give here the nonvanishing components of the stress tensor~:
\bea
T_t{}^t &=& \frac{1}{2} f'^2+\frac{\lambda ^2 f^2}{2K^2}+\frac{\lambda  n f^2
M}{L^2}+\frac{\lambda ^2 f^2 M^2}{2 L^2}+\frac{n^2 f^2}{2 L^2}-\frac{\lambda 
n f^2 M W^2}{N^2}-\frac{\lambda ^2 f^2   M v W}{N^2}-\frac{\lambda ^2 f^2
M^2 W^2}{2 N^2}-\frac{n^2 f^2 W^2}{2 N^2}\nonumber\\
&&-\frac{\lambda  n f^2 v
W}{N^2}-\frac{\lambda ^2 f^2 v^2}{2N^2}+\frac{\omega ^2
   f^2}{2 N^2}+V(f)\nonumber\\
T_r{}^r &=& -\frac{1}{2} f'^2+\frac{\lambda ^2 f^2}{2
K^2}+\frac{\lambda  n f^2 M}{L^2}+\frac{\lambda ^2
f^2 M^2}{2 L^2}+\frac{n^2 f^2}{2 L^2}-\frac{\lambda  n f^2
M W^2}{N^2}-\frac{\lambda ^2 f^2M v W}{N^2}-\frac{\lambda ^2 f^2
M^2 W^2}{2 N^2}-\frac{\lambda  \omega  f^2M W}{N^2}\nonumber\\
&&-\frac{n^2 f^2 W^2}{2
N^2}-\frac{\lambda  n f^2 v W}{N^2}-\frac{n\omega  f^2 W}{N^2}-\frac{\lambda
^2 f^2 v^2}{2 N^2}-\frac{\lambda  \omega  f^2 v}{N^2}-\frac{\omega ^2 f^2}{2
N^2}+V(f)\nonumber\\
T_\varphi{}^\varphi &=& \frac{1}{2} f'^2+\frac{\lambda ^2 f^2}{2
K^2}+\frac{\lambda ^2 f^2 M^2}{2 L^2}-\frac{n^2
f^2}{2 L^2}-\frac{\lambda ^2 f^2 M vW}{N^2}-\frac{\lambda ^2 f^2 M^2 W^2}{2
   N^2}-\frac{\lambda  \omega  f^2 MW}{N^2}+\frac{n^2 f^2 W^2}{2
N^2}-\frac{\lambda^2 f^2 v^2}{2 N^2}\nonumber\\
&&-\frac{\lambda  \omega  f^2v}{N^2}-\frac{\omega ^2 f^2}{2N^2}+V(f)\nonumber\\
T_z{}^z &=& \frac{1}{2} f'^2-\frac{\lambda ^2 f^2}{2
K^2}-\frac{\lambda ^2 f^2 M^2}{2 L^2}+\frac{n^2
f^2}{2L^2}+\frac{\lambda ^2 f^2 M vW}{N^2}+\frac{\lambda ^2 f^2 M^2 W^2}{2
   N^2}-\frac{n^2 f^2 W^2}{2 N^2}-\frac{n \omega f^2
W}{N^2}\nonumber\\
&&+\frac{\lambda ^2 f^2 v^2}{2 N^2}-\frac{\omega ^2 f^2}{2
N^2}+V(f)\nonumber\\
T_t{}^\varphi &=& -\frac{\lambda  \omega  f^2 M}{L^2}-\frac{n
\omega  f^2}{L^2}+\frac{\lambda  \omega  f^2 M W^2}{N^2}+\frac{n \omega  f^2
W^2}{N^2}+\frac{\lambda  \omega  f^2 v W}{N^2}+\frac{\omega ^2 f^2
W}{N^2}\nonumber\\
T_t{}^z &=& -\frac{\lambda  \omega  f^2}{K^2}-\frac{n \omega  f^2
M}{L^2}-\frac{\lambda  \omega  f^2 M^2}{L^2}+\frac{n \omega  f^2 M
W^2}{N^2}+\frac{2 \lambda  \omega  f^2 M v W}{N^2}+\frac{\lambda  \omega  f^2
M^2W^2}{N^2}+\frac{\omega ^2 f^2 MW}{N^2}\nonumber\\
&&+\frac{n \omega  f^2
v W}{N^2}+\frac{\lambda  \omega  f^2v^2}{N^2}+\frac{\omega
   ^2 f^2 v}{N^2}\nonumber\\
T_\varphi{}^z &=& -\frac{\lambda  n f^2}{K^2}-\frac{n^2 f^2
M}{L^2}-\frac{\lambda  n f^2 M^2}{L^2}+\frac{n^2 f^2 M W^2}{N^2}+\frac{2
\lambda  n f^2M v W}{N^2}+\frac{\lambda  n f^2 M^2 W^2}{N^2}+\frac{n
\omega f^2 MW}{N^2}\nonumber\\
&&+\frac{n^2 f^2 v W}{N^2}+\frac{\lambda  n f^2
v^2}{N^2}+\frac{n
\omega  f^2 v}{N^2}.
\eea

It can easily be shown that the seven equations are not
independant. Indeed, if one solves the off diagonal Einstein equations for
$W'',v'',M''$, injects the solution in the $r-r$ component of the Einstein
equation and further derives the resulting equation, then this equation is
indentically zero when the other equations are satisfied. In other words, the
derivative of the $r-r$ equation is a linear combination of the other
components.

\vskip 1.cm
\noindent{\bf Acknowledgments} We gratefully acknowledge G. Cl\'ement for pointing Ref. \cite{clement} to our attention.
\newpage

\end{document}